\documentclass[12pt,english]{article}
\pdfoutput=1
\usepackage[T1]{fontenc}
\usepackage{amssymb}
\usepackage{amsmath}
\usepackage{cite}
\usepackage{appendix}
\usepackage{epsfig}
\usepackage[unicode=true,
 bookmarks=true,bookmarksnumbered=false,bookmarksopen=false,
 breaklinks=false,pdfborder={0 0 1},backref=false,colorlinks=true]
 {hyperref}
\hypersetup{pdftitle={Entanglement Wedge Cross Section Inequalities from Replicated Geometries},
 pdfauthor={Ning Bao, Aidan Chatwin-Davies, Grant N. Remmen},
 citecolor=black,linkcolor=black,urlcolor=black}
\usepackage{breakurl}
\usepackage[hang,flushmargin]{footmisc}
\usepackage{breakurl}
\usepackage{color}
\usepackage{float}
\makeatletter
\def\simgt{\mathrel{\lower2.5pt\vbox{\lineskip=0pt\baselineskip=0pt
           \hbox{$>$}\hbox{$\sim$}}}}
\def\simlt{\mathrel{\lower2.5pt\vbox{\lineskip=0pt\baselineskip=0pt
           \hbox{$<$}\hbox{$\sim$}}}}
\makeatother

\newcommand{\Eq}[1]{Eq.~(\ref{#1})}

\newcommand{\Sec}[1]{Sec.~\ref{#1}}

\newcommand{\Fig}[1]{Fig.~\ref{#1}}

\newcommand{\Refs}[1]{Refs.~\cite{#1}}

\newcommand{\Hil}{\mathcal{H}}

\newcommand{\ket}[1]{| #1 \rangle}

\newcommand{\ketbra}[2]{|#1\rangle \langle #2 |}

\newcommand{\email}[1]{\href{mailto:#1}{\nolinkurl{#1}}}

\DeclareMathOperator{\Tr}{Tr}

\usepackage{xcolor}

\definecolor{purple}{rgb}{0.5,0,0.5}

\makeatletter

\def\lambdabar{\protect\@lambdabar}
\def\@lambdabar{%
\relax
\bgroup
\def\@tempa{\hbox{\raise.73\ht0
\hbox to0pt{\kern.25\wd0\vrule width.5\wd0
height.1pt depth.1pt\hss}\box0}}%
\mathchoice{\setbox0\hbox{$\displaystyle\lambda$}\@tempa}%
{\setbox0\hbox{$\textstyle\lambda$}\@tempa}%
{\setbox0\hbox{$\scriptstyle\lambda$}\@tempa}%
{\setbox0\hbox{$\scriptscriptstyle\lambda$}\@tempa}%
\egroup
}

\makeatother

\begin{document}

\interfootnotelinepenalty=10000
\baselineskip=18pt

\hfill

\vspace{2cm}
\thispagestyle{empty}
\begin{center}
{\LARGE \bf
Entanglement Wedge Cross Section \\ Inequalities from Replicated Geometries
}\\
\bigskip\vspace{1cm}{
{\large Ning Bao,${}^{a}$ Aidan Chatwin-Davies,$^{b,c}$ and Grant N. Remmen${}^{d,e}$}
} \\[7mm]
 {\it ${}^a$Computational Science Initiative, Brookhaven National Laboratory, Upton, New York, 11973 \\[1.5 mm]
  ${}^b$Department of Physics and Astronomy,
     University of British Columbia,\\[-1mm]6224 Agricultural Road, Vancouver, BC, V6T 1Z1\\[1.5mm]
     ${}^c$KU Leuven, Institute for Theoretical Physics,
 Celestijnenlaan 200D, B-3001 Leuven, Belgium}\\[1.5 mm]
 ${}^d${\it Kavli Institute for Theoretical Physics, University of California, Santa Barbara, CA 93106} \\[1.5mm]
 ${}^e${\it Department of Physics, University of California, Santa Barbara, CA 93106}
 \let\thefootnote\relax\footnote{\noindent e-mail: \email{ningbao75@gmail.com}, \email{achatwin@phas.ubc.ca}, \email{remmen@kitp.ucsb.edu}} \\
 \end{center}
\bigskip
\centerline{\large\bf Abstract}
\begin{quote} \small
    We generalize the constructions for the multipartite reflected entropy in order to construct spacetimes capable of representing multipartite entanglement wedge cross sections of differing party number as Ryu-Takayanagi surfaces on a single replicated geometry. We devise a general algorithm for such constructions for arbitrary party number and demonstrate how such methods can be used to derive novel inequalities constraining mulipartite entanglement wedge cross sections.
\end{quote}

\setcounter{footnote}{0}

\newpage
\tableofcontents
\newpage
    
\section{Introduction}
The study of properties of entanglement has had a profound impact on the study of AdS/CFT, with its first and perhaps most important result being the Ryu-Takayanagi (RT) formula \cite{Ryu_2006} relating areas of boundary-homologous minimal surfaces in the gravity theory to entanglement entropies of subregions in the boundary CFT. Recently, the study of bulk minimal surfaces and their boundary dual quantities has generalized away from the boundary-homologous requirement to the \textit{entanglement wedge cross section}~\cite{Umemoto_2018, Nguyen_2018}.
Conjectures for the boundary dual of this so-called $E_W$ surface include entanglement of purification \cite{Umemoto_2018, Nguyen_2018}, logarithmic negativity \cite{Kudler_Flam_2019}, and odd entropy \cite{Tamaoka_2019}. In the context of the entanglement of purification, the conjectured duality has been generalized to multipartite and conditional cases and tested in a number of ways; see for example \cite{bao2018holographic, Umemoto2_2018, bao2019conditional, harper2019bit, bao2019entanglement, bao2019towards, Akers_2020}.

A further conjectured dual to the entanglement wedge cross section is the reflected entropy,\footnote{An intriguing possibility is that all of the correspondences proposed thus far coincide in an appropriate limit (e.g., as $G_N \rightarrow 0$) or for an appropriate class of states.} first proposed in Ref.~\cite{dutta2019canonical}.\footnote{This proposal is natural, in that the doubled spacetime automatically includes a surface of twice the area of the entanglement wedge cross section as a RT surface of the double of the boundary subregion, and the doubling scheme is further highly suggestive of the so-called ``canonical purification/double'' of a quantum state. The correspondence is further supported by a replica calculation performed in Ref.~\cite{dutta2019canonical}.} This proposal relates the area of the entanglement wedge cross section to the area of a RT surface in a ``doubled'' geometry that holographically corresponds to a canonical purification of the original entanglement wedge.
Consequently, the area of the $E_W$ surface is given by the entanglement entropy of a boundary subregion in the canonically-purified state.
In Refs.~\cite{bao2019multipartite,Harper:2020wad, Chu_2020}, this proposal was generalized to the multipartite case.  
In particular, in all three generalizations, the multipartite entanglement wedge cross section maps onto a RT surface in a replicated geometry that consists of copies of the entanglement wedge.
The number of copies needed depends on number of parties (i.e., boundary subregions) in the original wedge.

This replication will be the central topic studied in this work. In the previous constructions, the focus was on generating a single replicated geometry on which a single, specific multipartite $E_W$ surface could be converted into (a fraction of) a RT surface. This naturally leads to the question of whether multiple different $E_W$ surfaces corresponding to different multipartite reflected entropies can all be converted into distinct RT surfaces on a single replicated geometry. In this case, as the replicated geometry would obey the holographic entanglement entropy inequalities studied in Refs.~\cite{bao2015holographic,Hayden_2013,avis2021foundations,cuenca2019holographic}, one would obtain potentially novel inequalities relating the $E_W$ surface areas of the original geometry.

In this work, we answer this question in the affirmative.
Beginning with an entanglement wedge for some fixed set of boundary subregions, one can construct a larger manifold on which certain RT surfaces correspond to $E_W$ surfaces for different numbers of boundary parties.
The larger manifold is constructed out of copies of the original wedge and its CPT conjugate, in a manner we will make explicit.
This construction allows us to prove new, nontrivial inequalities relating $E_W$ surfaces of different party numbers, e.g., Eqs.~\eqref{eq:SAEW}, \eqref{eq:SSAEW}, and \eqref{eq:MMIEW}.

The organization of this paper is as follows. In Sec.~\ref{sec:background}, we review the relevant background on multipartite entanglement wedge cross sections and the replication techniques referenced above. In Sec.~\ref{sec:example}, we work through an explicit construction of a replicated geometry that converts bipartite, tripartite, and four-partite $E_W$ surfaces into RT surfaces. Using our construction, we are able to demonstrate new inequalities---among $E_W$ surfaces of different party number---that are derivable from holographic entanglement entropy inequalities on the replicated geometry. In Sec.~\ref{sec:general}, we generalize our construction to make it applicable to arbitrary numbers of parties and spacetime dimensions. Finally, in Sec.~\ref{sec:discussion} we conclude with connections to existing replicated geometry methods and comments on future research directions. 

\section{Background and definitions}
\label{sec:background}

Let $\mathcal{M}$ be an asymptotically locally AdS (alAdS) manifold in $d+1$ spacetime dimensions that is holographically dual to a state of a CFT in $d$ spacetime dimensions, which we can think of as existing at the boundary of the bulk alAdS spacetime, $\partial \mathcal{M}$.
For simplicity, suppose that $\mathcal{M}$ is static or that it possesses a maximal spacelike slice that is a moment of time reflection symmetry.
Let $\Sigma$ denote this slice (or choose a maximal spacelike slice if $\mathcal{M}$ is static) and let $A \subseteq \partial \Sigma$ be a closed subregion on the boundary of $\mathcal{M}$.
In this setting, the RT formula gives a simple expression for the leading contribution to the von Neumann entropy of the CFT's reduced state on $A$ \cite{Ryu_2006}.
Let $m(A)$ denote the smallest-area $(d-1)$-dimensional minimal surface contained in $\Sigma$ that is homologous to $A$ (or one such surface if there are several with the same minimum area).
Then the RT formula reads
\begin{equation} \label{eq:rt}
    S(\rho_A) = \frac{|m(A)|}{4 G_N} + \text{(subleading)},
\end{equation}
where we use $| \cdot |$ to denote surface area.
Given $A$ and $m(A)$, we may define the \emph{entanglement wedge} of $A$, denoted $W_A$, as the ($d$-dimensional) interior of $A \cup m(A)$ on $\Sigma$ in this static or time-symmetric setting.\footnote{Note that our entanglement wedge should be denoted as the {\it homology region} corresponding to our boundary subregions, as our entanglement wedge is, strictly speaking, the intersection of the entanglement wedge proper with the given time slice.}

\begin{figure}[t]
    \centering
    \includegraphics[width=15cm]{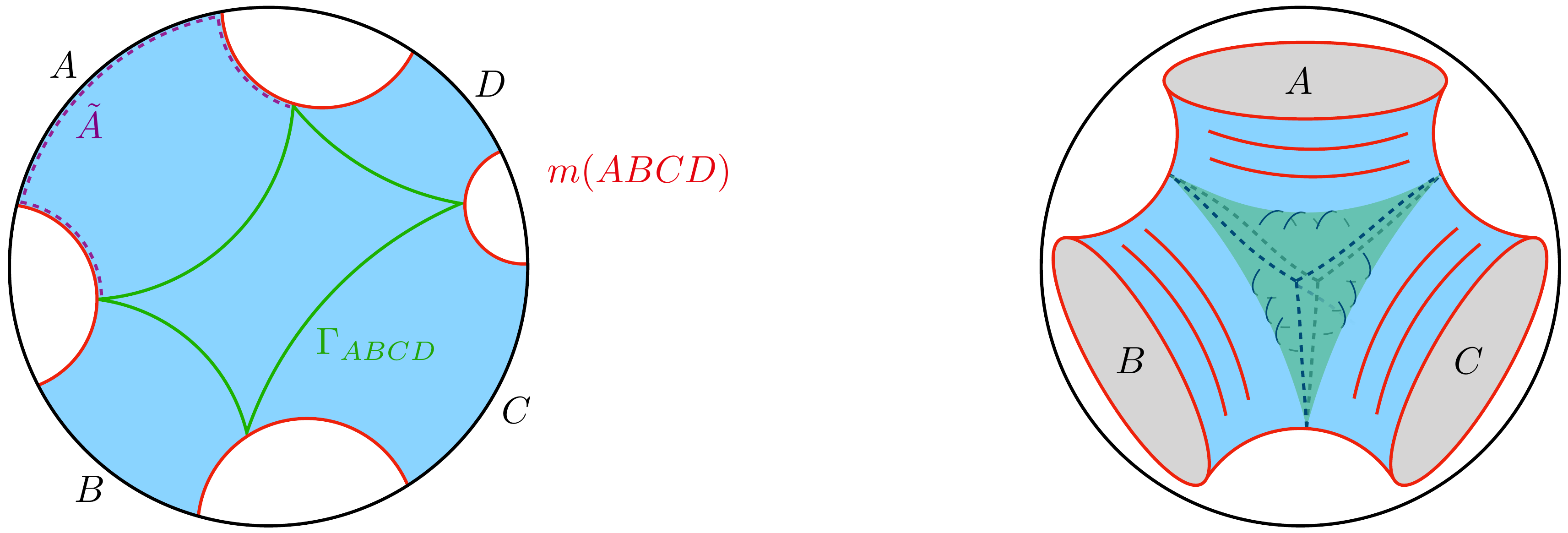}
    \caption{The Cauchy surface $\Sigma$ with entanglement wedge (blue) bounded by the given RT surface (red) and containing the multipartite entanglement wedge cross section (green) for: [Left] $n = 4$ boundary regions and $d=2$, with the part of the partition of $\partial W_{ABCD}$ that contains $A$, denoted $\tilde A$, additionally shown in dashed purple. [Right] $n=3$ boundary regions $A$, $B$, and $C$ in a CFT of spacetime dimension $d=3$.}
    \label{fig:wedge_examples}
\end{figure}

Now suppose that $A$ is the union of $n \geq 2$ nonintersecting subregions:
\begin{equation}
    A = A_1 \cup \cdots \cup A_n, \qquad A_i \cap A_j = \varnothing ~ ~ \text{when} ~ ~ i \neq j.
\end{equation}
Intuitively speaking, the \emph{$n$-partite entanglement wedge cross section}, which we denote by $\Gamma_{A_1 \cdots A_n}$, is the smallest-area $(d-1)$-dimensional minimal surface that is anchored to $m(A)$ and that partitions off $n$ subsets of $W_A$, each of which is connected to a single boundary component $A_i$.
In more precise language, let $\tilde A_1, \dots, \tilde A_n$ be a partition of $A\, \cup\, m(A)$ such that $\cup_{i=1}^n \tilde A_i = A \,\cup \,m(A)$, $A_i \subseteq \tilde A_i$, and any pair of distinct regions $\tilde A_i$ and $\tilde A_j$ only possibly meet at their boundaries (in other words, $\tilde a_i \cap \tilde a_j = \varnothing$ for any open subsets $\tilde a_i \subset \tilde A_i$ and $\tilde a_j \subset \tilde A_j$ whenever $i \neq j$).
Then, minimizing over all choices of partitions $\tilde A_1, \dots, \tilde A_n$, $\Gamma_{A_1 \cdots A_n}$ is the smallest-area $(d-1)$-dimensional minimal surface contained within $W_A$ such that $\Gamma_{A_1 \cdots A_n} \cap m(A) = \cup_{i=1}^n \partial \tilde{A}_i$ and that is homologous to $A \cup m(A)$.
A couple of examples are illustrated in \Fig{fig:wedge_examples}.
Also notice that $\Gamma_{A_1 \cdots A_n} = \varnothing$ is a possibility, occurring when the entanglement wedge consists of $n$ disconnected components.

Given $\Gamma_{A_1 \cdots A_n}$, its area defines the quantity $E_W(A_1 : {\cdots} : A_n)$ in analogy with the RT formula~\eqref{eq:rt},
\begin{equation}
    E_W(A_1 : \cdots : A_n) = \frac{|\Gamma_{A_1 \cdots A_n}|}{4 G_N}.
\end{equation}
In the case of two boundary regions, there is substantial evidence \cite{dutta2019canonical} that, to leading order,
\begin{equation} \label{eq:EWSR}
    E_W(A:B) = \frac{1}{2} S_R(A:B),
\end{equation}
where $S_R(A:B)$ is the \emph{reflected entropy} of the reduced state on $AB$.
For finite-dimensional Hilbert spaces\footnote{See Ref.~\cite{dutta2019canonical} for a careful discussion of the continuum CFT case.} $\Hil_A$ and $\Hil_B$, a straightforward definition of $S_R$ is as follows:
Let $\rho_{AB}$ be a density matrix in $\mathcal{L}(\Hil_{AB})$, and choose orthonormal bases of $\Hil_A$ and $\Hil_B$ so that we can write
\begin{equation}
    \rho_{AB} = \sum_{ii'jj'} \rho_{ii'jj'} \, \ketbra{i}{i'}_A \otimes \ketbra{j}{j'}_B
\end{equation}
for a collection of matrix elements $\rho_{ii'jj'}$.
The canonical purification of $\rho_{AB}$ is a pure state on a doubled Hilbert space $\Hil_{AA'BB'}$, with $\Hil_{A'} \cong \Hil_{A}$ and $\Hil_{B'} \cong \Hil_{B}$, given by
\begin{equation}
    \ket{\Psi} = \sum_{ii'jj'} \sqrt{\rho_{ii'jj'}} \, \ket{i}_A \ket{i'}_{A'} \ket{j}_B \ket{j'}_{B'} .
\end{equation}
The state $\ket{\Psi}$ has the property that $\rho_{AB} = \Tr_{A'B'} \ketbra{\Psi}{\Psi}$.
The reflected entropy is then defined as $S_R(A:B) = S(\rho_{AA'})$ or, written in a more symmetric way,
\begin{equation}
    S_R(A:B) = \frac{1}{2} \left[ S(\rho_{AA'}) + S(\rho_{BB'}) \right],
\end{equation}
where $\rho_{AA'} = \Tr_{BB'} \ketbra{\Psi}{\Psi}$ and $\rho_{BB'} = \Tr_{AA'} \ketbra{\Psi}{\Psi}$.

\begin{figure}[t]
    \centering
    \includegraphics[width=15cm]{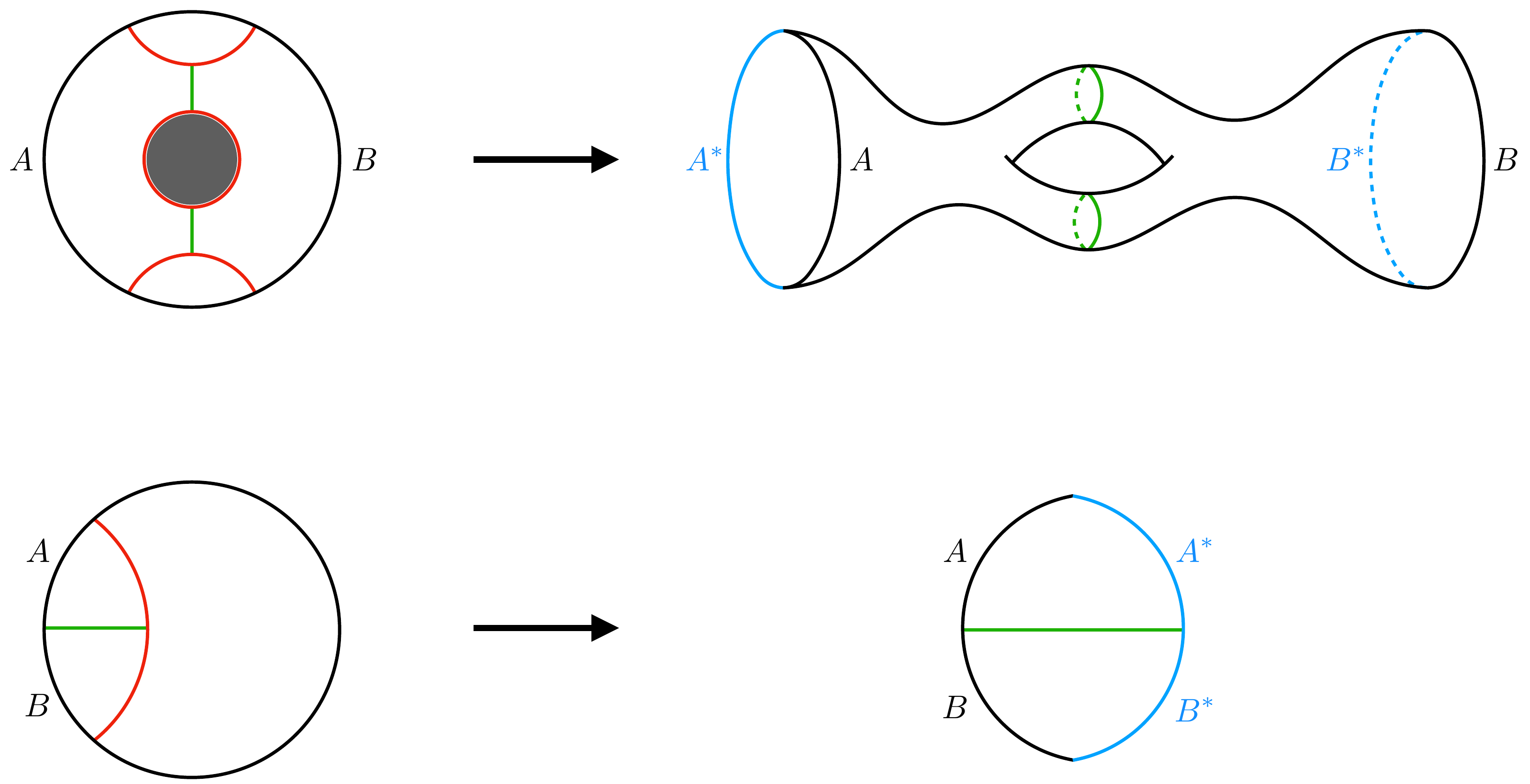}
    \caption{Doubling $W_{AB}$ and gluing along $m(AB)$ produces the geometry that is holographically dual to the canonical purification of $\rho_{AB}$. [Top] Two disjoint boundary regions for a state that contains a black hole in the bulk. $\Gamma_{AB}$ and its CPT conjugate join up to form two disjoint closed loops that are together homologous to $AA^*$. [Bottom] Two boundary regions that share a boundary. $\Gamma_{AB}$ and its CPT conjugate join up to form a boundary-anchored geodesic that is homologous to $AA^*$.}
    \label{fig:gluing_examples}
\end{figure}

Given a reduced CFT state $\rho_{AB}$ that has an entanglement wedge $W_{AB}$, it has been argued that the holographic dual of its canonical purification is constructed by taking a CPT-conjugate copy of the wedge, $W_{A^*B^*}$, and gluing it to $W_{AB}$ along $m(AB)$ and $m(A^*B^*)$, with the canonical purification itself being supported on the CPT-doubled space $AA^*BB^*$ \cite{dutta2019canonical}.
When $A \cup B \neq \partial \Sigma$ and $A$ and $B$ do not share any portions of their boundaries (and if $\Gamma_{AB}$ is nonempty), it follows that the components of $\Gamma_{AB}$ and $\Gamma_{A^*B^*}$ join up to form closed minimal surfaces without boundary, since $\partial \Gamma_{AB}$ lies on $m(AB)$.
If $A \cup B$ makes up the entire conformal boundary or the boundaries of $A$ and $B$ touch, then $\Gamma_{AB}$ and $\Gamma_{A^*B^*}$ can join to form boundary-anchored minimal surfaces, or they may already individually be surfaces without boundary in the bulk.
In all cases, it therefore follows that the two cross sections $\Gamma_{AB}$ and $\Gamma_{A^* B^*}$ join up to form minimal surfaces that are homologous to $AA^*$ and $BB^*$, i.e., RT surfaces, and so it is pictorially clear that \Eq{eq:EWSR} holds.
(See \Fig{fig:gluing_examples} for examples.)

The key idea from the holographic construction above that we wish to focus on is that the bulk-anchored cross section $\Gamma_{AB}$ is realized as a conventional boundary-homologous minimal surface on a larger replicated manifold.
Alternatively, according to the holographic dictionary, $E_W(A:B)$ is realized as a conventional holographic entanglement entropy.
Similar constructions, such as those described in \Refs{bao2019multipartite,Harper:2020wad}, map multipartite cross sections to minimal surfaces in even larger manifolds.
However, these constructions are only ever guaranteed to compute $E_W(A_1 : A_2 : \cdots : A_n)$ alone as an entanglement entropy.
Our goal, which we take up in the following section, will be to construct a replicated manifold on which a substantial family of $m$-partite $E_W$ surfaces are simultaneously realized as entanglement entropies for all $1 < m \leq n$.

Although it is not directly relevant to the construction that we will pursue, for the sake of completeness we note that $E_W(A_1 : \cdots : A_n)$ is similarly conjectured to be proportional to a multipartite version of reflected entropy in the boundary CFT~\cite{bao2019multipartite}.
However, less work has been done to precisely establish the correspondence as compared to the bipartite case.

While we focus on the static or time reflection-symmetric cases here, in principle there is no obstruction to working with a fully time-dependent spacetime $\mathcal{M}$.
In such a setting, the minimal RT surface is replaced by the extremal Hubeny-Rangamani-Takayanagi (HRT) surface \cite{Hubeny_2007}, and the entanglement wedge is a $(d+1)$-dimensional bulk spacetime domain of dependence~\cite{Headrick_2014}.
Nevertheless, within the covariant construction, one can still pick out a preferred bulk Cauchy surface that contains the HRT surface (a maximin surface \cite{Wall_2014}) so that cross sections may be computed and the gluing may be carried out according to the methods outlined above~\cite{bao2019multipartite}.

\section{Construction and inequalities for four parties}
\label{sec:example}

\subsection{Four-party construction}\label{sec:fourpartyconstruction}

\begin{figure}[H]
\begin{center}
\includegraphics[width=14cm]{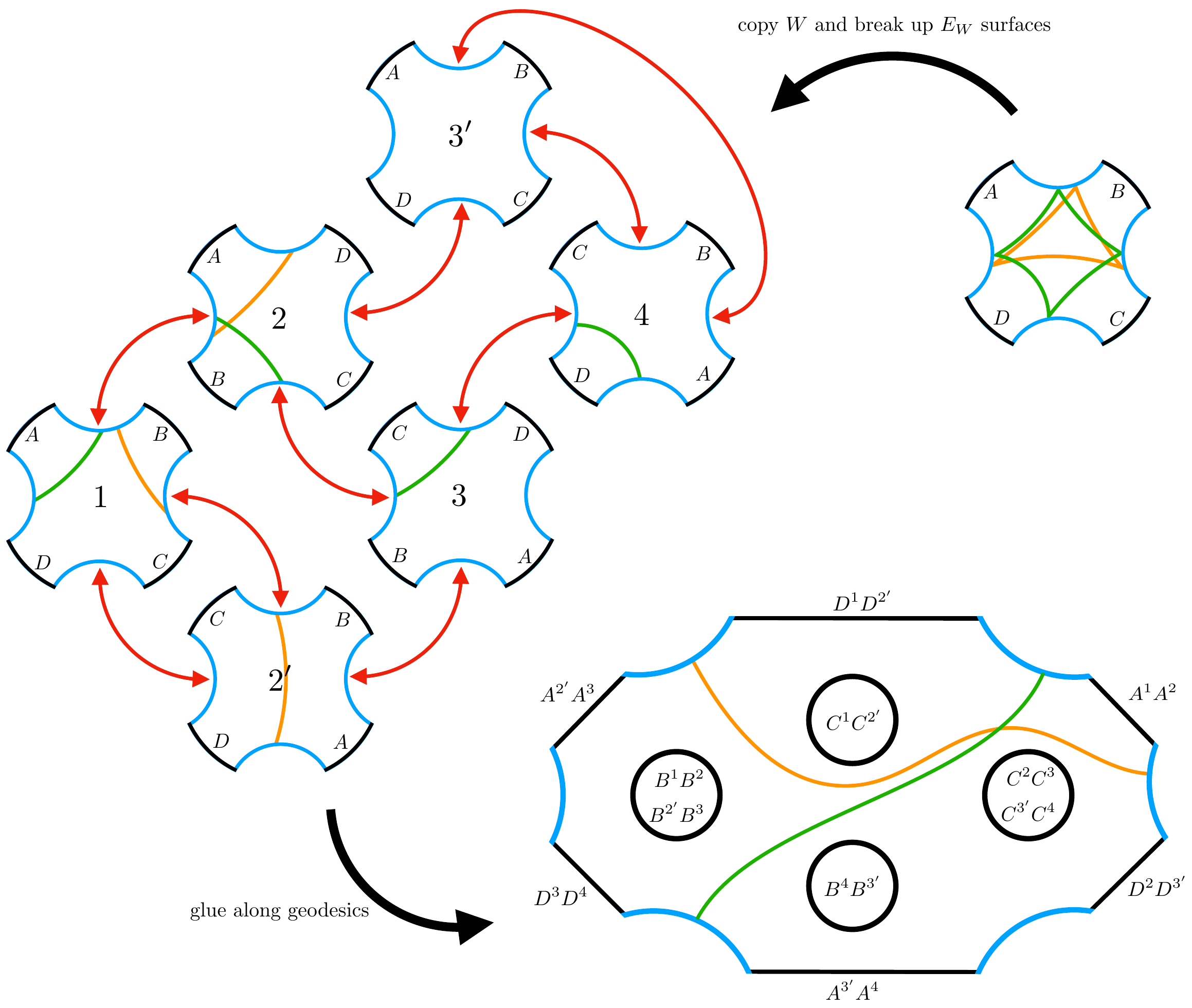}
\end{center}
\vspace{-0.5cm}
\caption{The entanglement wedge of $A, B, C, D$ is replicated and glued to form a manifold $M_0$ on which multipartite entanglement wedge cross sections are mapped to $(A,D)$-anchored geodesics. For illustration, the surface corresponding to $E_W(A:B:C:D)$ is mapped to a geodesic that subtends $A^1A^2A^{3'}A^4B^{3'}B^4C^2C^3C^{3'}C^4D^2D^{3'}$ (shown in green), and the surface corresponding to $E_W(A:B:CD)$ is mapped to a geodesic that subtends $A^1A^2C^1C^{2'}D^1D^{2'}$ (shown in orange).}
\label{fig:fourpartyconstruction}
\end{figure}

Before presenting our general algorithm for extracting inequalities among $E_W$ surfaces of general party number, it will be illuminating to first consider a concrete example.
Let us take a CFT in $1+1$ dimensions, dual to an asymptotically ${\rm AdS}_3$ bulk, and identify four subregions on a spacelike slice of the boundary, which we label as $A$, $B$, $C$, and $D$, defining an entanglement wedge $W$ depicted in \Fig{fig:fourpartyconstruction}. 
We will require that the four subregions be chosen such that $W$ is in the fully-connected phase, and for simplicity let us further assume that $A$, $B$, $C$, and $D$ are each comprised of a single, simply-connected interval on the boundary.

Within $W$, we can identify various multiparty entanglement wedge surfaces, e.g., bipartite surfaces like $E_W(AB:CD)$, tripartite surfaces like $E_W(AB:C:D)$, and the four-partite surface $E_W(A:B:C:D)$.
We would like to reinterpret the areas of these various $E_W$ surfaces as a holographic entanglement entropy, which will allow us to construct new inequalities among the  entanglement wedge cross sections.
To do so, let us replicate our original entanglement wedge $W$, with the ultimate goal of constructing a larger spacetime in which the cross sections are transformed into RT surfaces.
The Israel junction conditions imply that we can glue together spacetimes along extremal codimension-one surfaces in general spacetime dimension~\cite{Engelhardt_2019,dutta2019canonical}.
For the static, $(2+1)$-dimensional case at hand, this means that we can form a spacelike slice of a larger geometry by gluing together copies of spacelike regions along the same geodesic in the two copies.

Let us take four copies of $W$, labeled $W^1$ through $W^4$, as well as two additional copies labeled $W^{2'}$ and $W^{3'}$.
To allow for a bulk theory containing fermions or charged matter, let the fields in even-numbered copies ($W^2$, $W^4$, and $W^{2'}$) be CPT conjugates of the odd-numbered copies.

We write the bulk geodesic that connects an endpoint of $A$ to the adjacent one of $B$ as $(A,B)$, etc., and will write an identification of a given geodesic $(E,F)$ in wedge $U$ with the corresponding one in wedge $V$ as $U\stackrel{(E,F)}{\longleftrightarrow} V$.
From our six copies of $W$, let us build a single geometry by making the following identifications, depicted in \Fig{fig:fourpartyconstruction}:
\begin{equation}
\begin{aligned}
W^1 \stackrel{(A,B)}{\longleftrightarrow} W^2 \\
W^2 \stackrel{(B,C)}{\longleftrightarrow} W^3 \\
W^3 \stackrel{(C,D)}{\longleftrightarrow} W^4,
\end{aligned}
\end{equation}

along with 
\begin{equation}
\begin{aligned}
W^{1} \stackrel{(B,C)}{\longleftrightarrow} W^{2'} \\
W^{1} \stackrel{(C,D)}{\longleftrightarrow} W^{2'} \\
W^{2'} \stackrel{(A,B)}{\longleftrightarrow} W^3 \\
W^{2} \stackrel{(C,D)}{\longleftrightarrow} W^{3'}\\
W^{3'} \stackrel{(A,B)}{\longleftrightarrow} W^4 \\
W^{3'} \stackrel{(B,C)}{\longleftrightarrow} W^4.
\end{aligned}
\end{equation}
The end product is a connected manifold, where one of the boundaries is formed by the disjoint union of $A^1 A^2$, $A^{2'} A^3$, $A^{3'} A^4$, $D^1 D^{2'}$, $D^2 D^{3'}$, and $D^3 D^4$ along with six copies of the geodesic segment $(D,A)$, where we write $A^i$ for the copy of $A$ in wedge $W^i$ and analogously for the other regions.

\begin{figure}[t]
\begin{center}
\includegraphics[width=\textwidth]{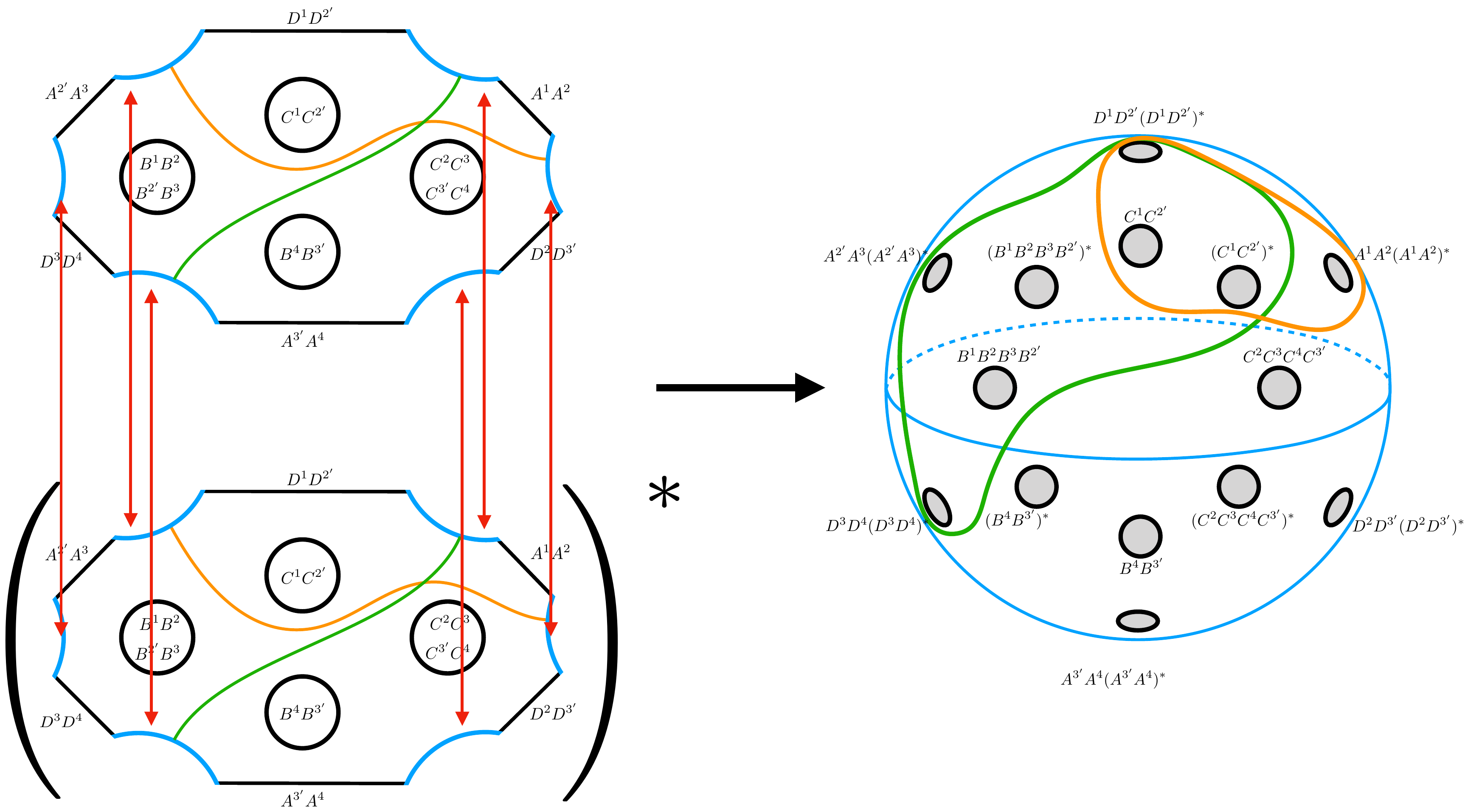}
\end{center}
\vspace{-0.5cm}
\caption{$M_0$ and its CPT conjugate $M_0^*$ are glued along the open $(A,D)$ geodesics to form the final manifold $M$ on which the geodesics that correspond to $E_W$ surfaces of the original entanglement wedge are closed loops.}
\label{fig:finalM}
\end{figure}

\begin{figure}[H]
\begin{center}
\includegraphics[width=12cm]{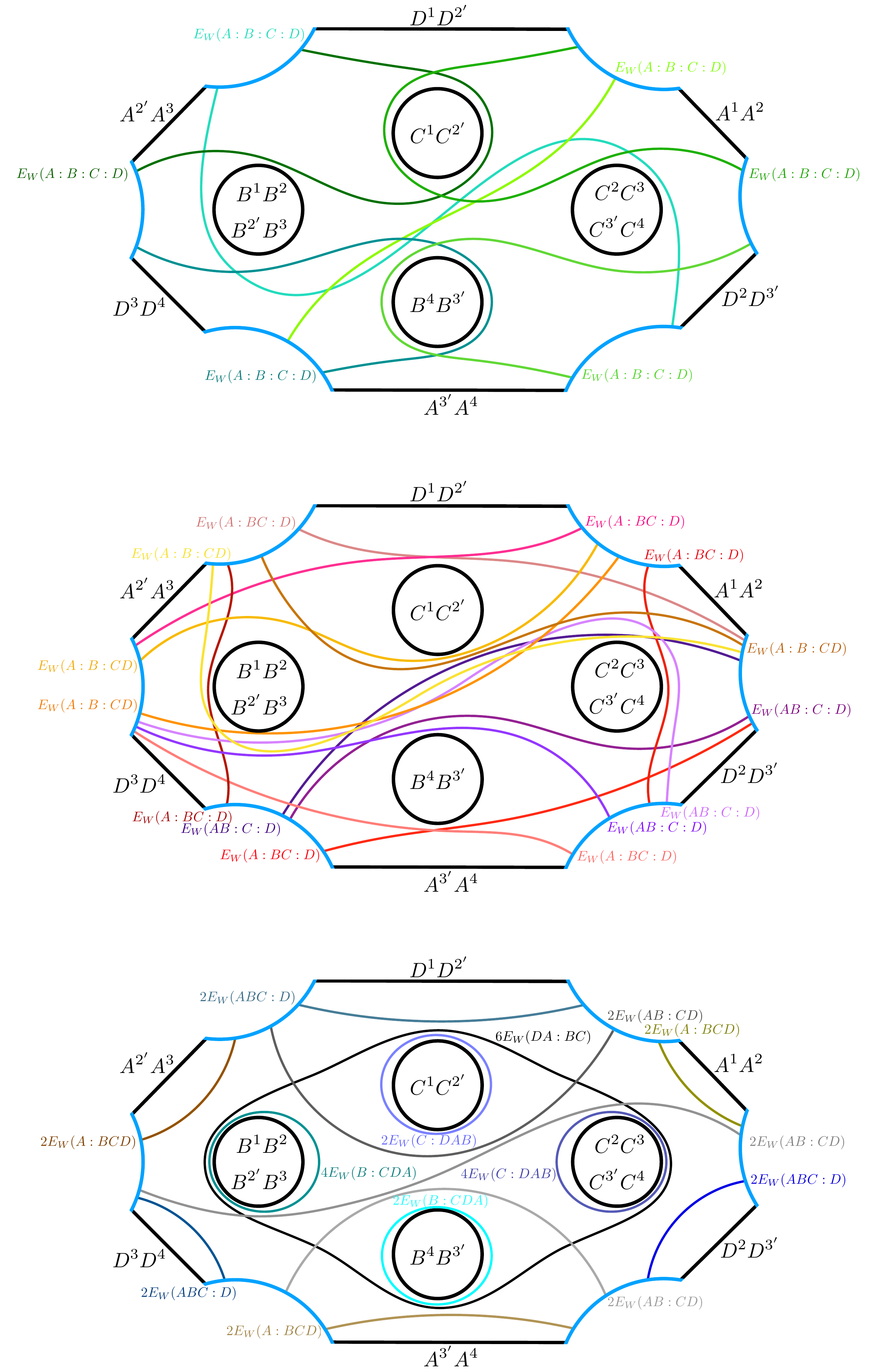}
\end{center}
\vspace{-0.5cm}
\caption{$(A,D)$-anchored geodesics on $M_0$ that correspond to all $E_W$ surfaces achievable with our construction, with surfaces of different types shown as shades of different colors. The (integer multiple of the) $E_W$ that each geodesic's length computes is indicated. [Top] The four-party $E_W(A:B:C:D)$ surfaces.  [Middle] Three-party surfaces. [Bottom] Two-party surfaces.
}
\label{fig:EWs}
\end{figure}

We construct our final spacetime by taking $M_0$ and its CPT conjugate $M_0^*$ and gluing them along all of the corresponding copies of the $(D,A)$ geodesic (the blue edges in the figures).
The result is a Cauchy surface for a spacetime $M$ whose only boundaries are closed loops made of copies of the original boundary subregions of $W$; see \Fig{fig:finalM}.
A consequence of this construction is that the surface corresponding to $E_W(A:B:C:D)$ in $W$ becomes, in the replicated-and-glued geometry $M_0$, a single geodesic anchored to two different copies of $(D,A)$ that passes through $W^1$, $W^2$, $W^3$, and $W^4$.
Other geodesics can be identified in $M_0$ that correspond to various bipartite and tripartite entanglement wedge cross sections; see \Fig{fig:EWs}.
In fact, $M_0$ contains RT surfaces corresponding to all two-, three-, and four-party $E_W$ surfaces that respect the ordering of the parties $A$, $B$, $C$, and $D$.
The correspondences between various RT and $E_W$ surfaces still hold in $M$ as they did in $M_0$.

The construction that we have proposed here can be thought of as a compromise between minimizing the number of copies and symmetry.
A minimum of four copies ($W^1$ through $W^4$) are needed to obtain the four-party $E_W$ surface as a RT surface.
We then introduced two additional copies ($W^{2'}$ and $W^{3'}$) in order to obtain the two- and three-party $E_W$ surfaces.
In principle, at least one RT surface for each two- and three-party $E_W$ would have been obtainable had we introduced only one of $W^{2'}$ or $W^{3'}$; however, using all of the copies as described above ensures that the boundary components of $M_0$ that consist of bulk geodesics are only made up of copies of the $(D,A)$ geodesic.
In particular, this structure lets us straightforwardly generalize the construction to higher party numbers, as we describe in \Sec{sec:general}.
We come back to the question of other possible gluing schemes in \Sec{sec:discussion}.

\subsection{Inequalities}
\label{sec:ineqs}

Let us use the four-party construction detailed in \Sec{sec:fourpartyconstruction} to derive new inequalities among $E_W$ surfaces.
These new inequalities are direct consequences of relationships between various entanglement entropies computed on the replicated geometry $M$.
Since $M$ is topologically a sphere with fourteen punctures, as one can see from \Fig{fig:finalM}, a complete characterization of all potential $E_W$ relations derivable from this construction would require implementation of the fourteen-party holographic entropy cone\footnote{The holographic entropy cone is defined as the conical region in the space of von Neumann entropies of all subsystems, whose faces are defined by entanglement inequalities satisfied either by all quantum states---e.g., SA and SSA---or holographic states, e.g., MMI \cite{Hayden_2013}.}, which would not be tractable~\cite{bao2015holographic,cuenca2019holographic}.
Instead, it will be useful---and will give a flavor of the power of this construction---to consider new $E_W$ inequalities that we can derive from three notable entanglement relations: subadditivity (SA), strong subadditivity (SSA), and monogamy of mutual information (MMI).
We will give a single example of each in turn; while this is by no means exhaustive, it will suffice to illustrate the types of inequalities generable using this method.

\begin{figure}[H]
\begin{center}
\includegraphics[width=12cm]{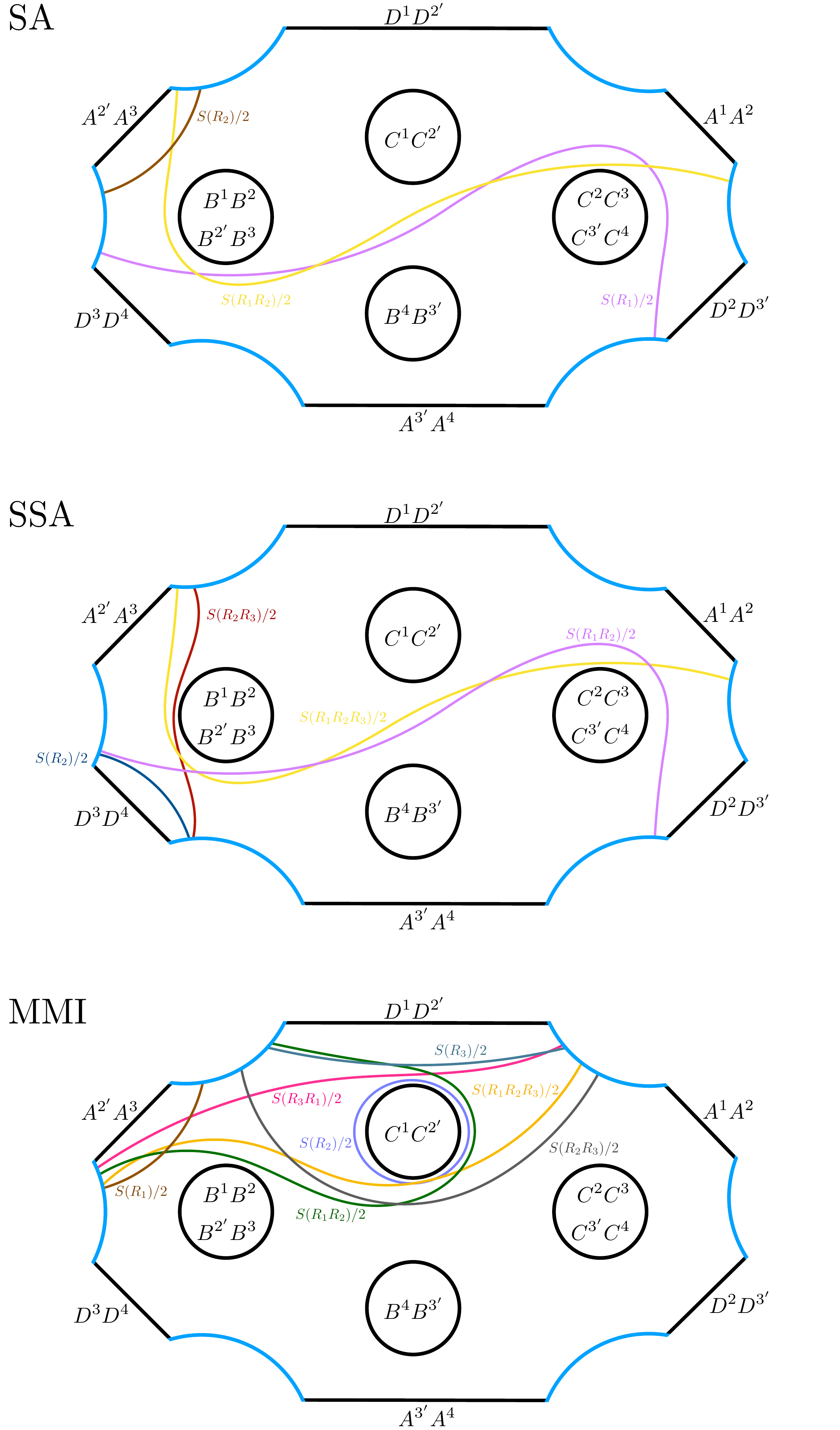}
\end{center}
\vspace{-0.5cm}
\caption{$E_W$ surfaces for the choices of $R_i$ in Eqs.~\eqref{eq:R12choice}, \eqref{eq:R123choice}, and \eqref{eq:R123choiceMMI} (prior to the final doubling of the manifold using the CPT conjugate), where we use SA, SSA, and MMI, respectively, to derive the inequalities in Eqs.~\eqref{eq:SAEW}, \eqref{eq:SSAEW}, and \eqref{eq:MMIEW}.
}
\label{fig:examples}
\end{figure}

First, let us consider SA:
\begin{equation}
S(R_1 R_2) \leq S(R_1) + S(R_2).\label{eq:SA}
\end{equation}
Identifying $R_1$ and $R_2$ with various choices of regions on the boundary of $M$, we obtain new inequalities upper-bounding the four-party $E_W$ surface in terms of various two- and three-party quantities.
Choosing 
\begin{equation}
\begin{aligned}
R_1 &= A^4 A^{3'}(A^4 A^{3'})^* D^3 D^4 (D^3 D^4)^* B^4 B^{3'} (B^4 B^{3'})^* C^2 C^3 C^4 C^{3'}(C^2 C^3 C^4 C^{3'})^* \\
R_2 &= A^3 A^{2'} (A^3 A^{2'})^*
\end{aligned}\label{eq:R12choice}
\end{equation}
and translating Eq.~\eqref{eq:SA} into $E_W$ surfaces (see \Fig{fig:examples}), we find
\begin{equation} \label{eq:SAEW}
\boxed{
E_W(A:B:C:D) \leq 2 E_W(A:BCD) + E_W(AB:C:D).
}
\end{equation}

Let us next consider how to obtain new $E_W$ relations from SSA:
\begin{equation}
S(R_1 R_2 R_3) + S(R_2) \leq S(R_1 R_2)+S(R_2 R_3).
\end{equation}
Taking 
\begin{equation}
\begin{aligned}
R_1 &= A^4 A^{3'}(A^4 A^{3'})^* B^4 B^{3'} (B^4 B^{3'})^* C^2 C^3 C^4 C^{3'}(C^2 C^3 C^4 C^{3'})^*\\
R_2 &= D^3 D^4 (D^3 D^4)^*\\
R_3 &= A^3 A^{2'}(A^3 A^{2'})^*,
\end{aligned}\label{eq:R123choice}
\end{equation}
we find a new relation among two-, three-, and four-party $E_W$ surfaces (see \Fig{fig:examples}):
\begin{equation}
\boxed{
E_W (A:B:C:D) + 2 E_W(ABC:D) \leq E_W(A:BC:D) + E_W(AB:C:D).
}\label{eq:SSAEW}
\end{equation}

Finally, let us turn to MMI.
Unlike SA or SSA, MMI is not satisfied for general quantum states, but holds holographically in the large-$N$ limit~\cite{Hayden_2013}.
The statement of MMI is
\begin{equation}
S(R_1) + S(R_2) + S(R_3) + S(R_1 R_2 R_3) - S(R_1 R_2) - S(R_2 R_3) - S(R_3 R_1)\leq 0.\label{eq:MMI}
\end{equation}
The left-hand side of \Eq{eq:MMI} equals $I_3 (R_1{:}R_2{:}R_3)$, the tripartite mutual information.
Taking $R_{1,2,3}$ as follows,
\begin{equation}
\begin{aligned}
R_1 &= A^{2'} A^3 (A^{2'}A^3)^* \\
R_2 &= C^1 C^{2'}  (C^1 C^{2'})^* \\
R_3 &= D^1 D^{2'} (D^1 D^{2'})^*, \\
\end{aligned}\label{eq:R123choiceMMI}
\end{equation}
implies another new inequality among two-, three-, and four-party $E_W$ quantities (see \Fig{fig:examples}):
\begin{equation}
\boxed{
\begin{aligned}
E_W(A{:}B{:}CD) + 2\left[E_W(A{:}BCD) + E_W(C{:}DAB) + E_W(D{:}ABC) \right] \\
\leq E_W(A{:}B{:}C{:}D) + E_W(A{:}BC
{:}D) + 2 E_W(AB{:}CD).
\end{aligned}
}\label{eq:MMIEW}
\end{equation}
Notably, while we proved an upper bound on the four-party $E_W$ surface via SSA in \Eq{eq:SSAEW}, we find a lower bound on the four-party $E_W$ in \Eq{eq:MMIEW}.

It is unknown whether the inequalities in Eqs.~\eqref{eq:SAEW}, \eqref{eq:SSAEW}, and \eqref{eq:MMIEW} are tight.
To establish tightness, one would need to find holographic states saturating these inequalities, a task we leave to future study.
A further open question is to what degree these inequalities are true only for holographic states or for quantum states more generally.
If these inequalities are true only for holographic states, they provide yet more constraints that quantum states must satisfy in order to potentially possess a semiclassical gravity dual.
These novel inequalities are of particular interest because they mix party number, e.g., bounding the four-party $E_W$ surface in terms of its two- and three-party counterparts.

\newpage

\section{General construction for an arbitrary number of parties}
\label{sec:general}

In this section, we generalize our construction for $n \geq 2$ parties.
Doing so would allow us, in principle, to derive yet more inequalities, for greater numbers of parties, analogous to those in Sec.~\ref{sec:example}.
Such a process would ultimately form a cone among the holographically allowed $E_W$ quantities (in analogy with the holographic entropy cone).
While for brevity we will not list more inequalities in this section, we will present the general construction for arbitrary party number.
The construction is an extension of the four-party case discussed in \Sec{sec:example}, and so its assembly has a similar intuitive motivation: introduce a minimal number of copies of the entanglement wedge so that $E_W$ surfaces for all party numbers are realized as RT surfaces, while ultimately maintaining a certain amount of geometric symmetry.

Let $A_1, A_2, \dots, A_n$ be an ordered list of nonintersecting, simply-connected boundary subregions with an entanglement wedge $W$ that is in the fully-connected phase.
The output of our construction will be a manifold that consists of copies of $W$ and its CPT conjugate $W^*$, glued together along minimal surfaces, on which RT surfaces correspond to $E_W$ surfaces in $W$.
The $E_W$ surfaces obtained in this way are the $m$-party cross sections for all partitions of the ordered list $A_1, A_2, \dots, A_n$ into $m$ groups for $2 \leq m \leq n$.
In other words, we obtain $E_W(\alpha_1 : \cdots : \alpha_m)$ for all partitions
\begin{equation} \label{eq:partition}
\begin{aligned}
    \alpha_1 &= A_1 A_2 \cdots A_{q_1} \\
    \alpha_2 &= A_{q_1 + 1} \cdots A_{q_2} \\
    \vdots \\
    \alpha_m &= A_{q_{m-1}+1} \cdots A_n
\end{aligned}
\end{equation}
where $1 \leq q_1 < q_2 < \cdots < q_{m-1} < n$.

\subsection{Construction for 2+1 dimensions}
\label{sec:2d}

We first consider the case where $W$ is two-dimensional.
Let us assume that $A_1, A_2, \dots, A_n$ are ordered sequentially around the boundary and, for now, assume that $n$ is even.
Denote the geodesic that connects an endpoint of $A_i$ to an endpoint of $A_{i+1}$ by $(i,i+1)$.
The algorithm for constructing the extended manifold $M_0$ is as follows:
\begin{enumerate}
    \item Take $n$ copies of $W$ that we label $W^1, W^2, \dots, W^n$, as well as $n-2$ more copies that we denote by $W^{2'}, \dots, W^{(n-1)'}$.
    Moreover, suppose that even-numbered copies are CPT conjugates of odd-numbered copies so that the following identifications are possible.
    \item Connect each copy $W^i$ to $W^{i-1}$ by making the identification
    \begin{equation*}
        W^{i} \stackrel{(i-1,i)}{\longleftrightarrow} W^{i-1}
    \end{equation*}
    as well as to $W^{i+1}$ by making the identification
    \begin{equation*}
        W^{i} \stackrel{(i,i+1)}{\longleftrightarrow} W^{i+1}
    \end{equation*}
    for each $2 \leq i \leq n-1$.
    $W^1$ and $W^n$ do not have these first and last identifications, respectively.
    \item Connect each primed copy ${W}^{i'}$ to the unprimed copy $W^{i+1}$ by gluing along
    \begin{equation*}
        \begin{aligned}
            W^{i'} &\stackrel{(1,2)}{\longleftrightarrow} W^{i+1} \\
            W^{i'} &\stackrel{(2,3)}{\longleftrightarrow} W^{i+1} \\
            &\;\;\;\; \vdots \\
            W^{i'} &\stackrel{(i-1,i)}{\longleftrightarrow} W^{i+1}
        \end{aligned}
    \end{equation*}
    as well as to the unprimed copy $W^{i-1}$ by gluing along 
    \begin{equation*}
        \begin{aligned}
            W^{i'} &\stackrel{(i,i+1)}{\longleftrightarrow} W^{i-1} \\
            W^{i'} &\stackrel{(i+1,i+2)}{\longleftrightarrow} W^{i-1} \\
            &\;\;\;\; \vdots \\
            W^{i'} &\stackrel{(n-1,n)}{\longleftrightarrow} W^{i-1}
        \end{aligned}
    \end{equation*}
     for each $2 \leq i \leq n-1$.
\end{enumerate}
The result is a connected manifold, $M_0$, that is topologically a disc with punctures.
The exterior boundary of $M_0$ is the disjoint union of $A^1_1 A^2_1$ and $A^{i'}_1 A^{i+1}_1$ for $2 \leq i \leq n-1$, $A^{j}_n A^{(j+1)'}_n$ for $1 \leq j \leq n-2$ and $A^{n-1}_n A^n_n$, as well as the geodesic boundary $(n,1)^q$ on each copy for $1 \leq q \leq n$ and $(n,1)^{i'}$ for $2 \leq i \leq n-1$.
As before, we use $A_k^q$ to denote the boundary subregion $A_k$ on copy $W^q$, and analogously we use $(i,i+1)^q$ to denote the geodesic $(i,i+1)$ on copy $W^q$.
The interior boundaries are closed loops that are the unions of two or four copies of the same boundary region $A_i$ for $2 \leq i \leq n-1$.

This particular construction guarantees that the original surface in $W$ corresponding to $E_W(A_1 : \cdots : A_n)$ becomes a geodesic in $M_0$ that is anchored to $(n,1)^1$ and $(n,1)^n$. 
This geodesic passes through each unprimed copy sequentially, and its segment on each tile $W^i$---that is, each copy of the original entanglement wedge---subtends $A_i^i$.
Upon doubling $M_0$ by gluing it to its CPT conjugate $M_0^*$ along all of the $(n,1)$ boundaries, this geodesic becomes a closed, boundary-homologous minimal curve, or in other words, a RT surface.
The choice of $n$ being even ensures that this gluing preserves parity consistently across the final manifold.

In fact, there are in total $2n-2$ such geodesics on $M_0$---one for each tile, primed or unprimed---that start and end on copies of $(n,1)$ and that correspond to the $n$-party $E_W$.
Begin by picking a starting tile, on which the geodesic's segment subtends $A_1$.
The geodesic then passes through $(1,2)$, and its next segment on the next tile subtends $A_2$, and so on until $n$ segments have been accumulated. 

\begin{figure}[t]
    \centering
    \includegraphics[width=14cm]{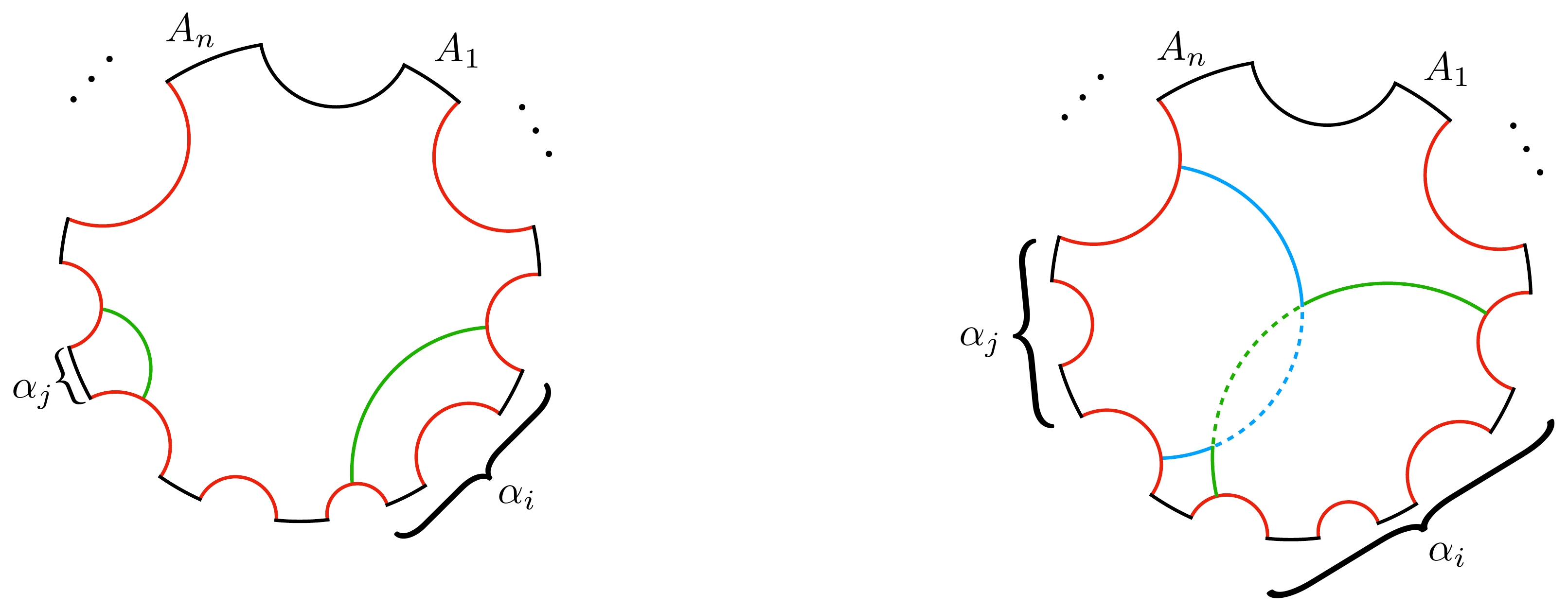}
    \vspace{-5mm}
    \caption{[Left] An example of segments of the surface corresponding to $E_W(\alpha_1 : \cdots : \alpha_m)$ on a single tile that subtend the collections of boundary subregions $\alpha_i$ and $\alpha_j$, with $i < j$. [Right] An inconsistent configuration; the blue and green segments cannot both be minimal if the dashed parts have different length. If they have exactly the same length, then we can redefine one of the segments by letting it run along the other's dashed component.}
    \label{fig:m-party}
\end{figure}

By extension, we see that any $m$-party $E_W$ surface for a partition of the form \eqref{eq:partition} will map onto a geodesic that starts and ends on a boundary geodesic $(n,1)$ (possibly on the same tile).
On the starting tile, the geodesic's segment subtends $\alpha_1$; on the next tile, its segment subtends $\alpha_2$; and so on up to $\alpha_m$.
The condition that the partition itself retain the ordering of the boundary subregions ensures that the geodesic never intersects itself, even if it visits the same tile multiple times.\footnote{It is technically possible, in a finely-tuned setup, that the geodesic could overlap with itself over some non-zero length, although this does not change the final conclusion.} See \Fig{fig:m-party} for illustration.
Therefore, after the final doubling, we see that these $m$-party $E_W$ surfaces end up corresponding to closed boundary-homologous geodesics.

\begin{figure}[t]
    \centering
    \includegraphics[width=6.6cm]{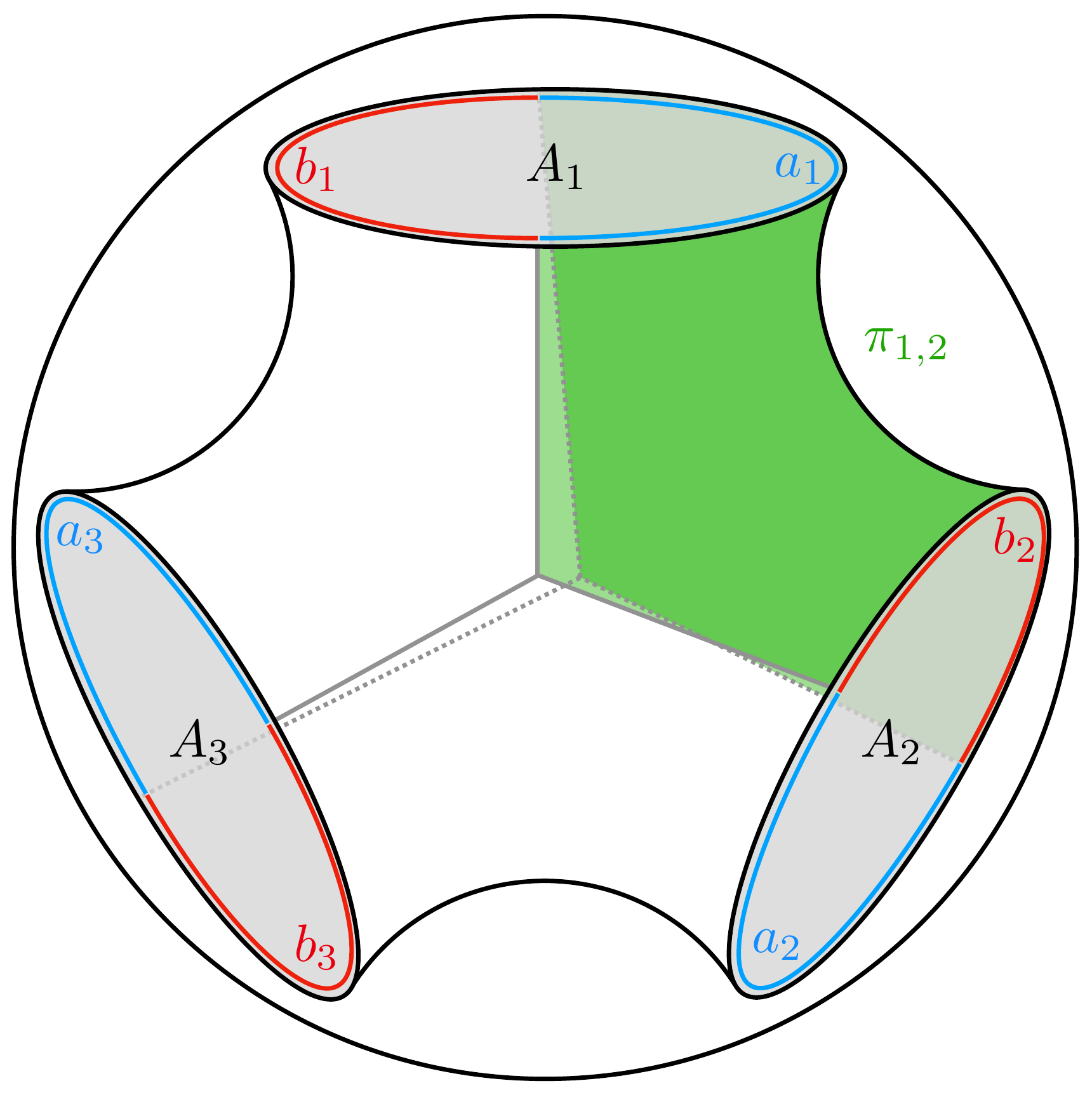}
    \vspace{-3mm}
    \caption{A three-dimensional entanglement wedge for three boundary regions in the fully-connected phase. Its bounding RT surface must be partitioned into components that connect sequential pairs of boundary subregions, analogously to how geodesics connect pairs of boundary subregions in two dimensions. For illustration, the component $\pi_{1,2}$ connecting $A_1$ to $A_2$ is shaded.}
    \label{fig:partitionD}
\end{figure}

We initially assumed that $n$ was even so that parity could be preserved across $M_0$ and $M_0^*$ when they are joined.
In the case where $n$ is odd, we can avoid any inconsistencies by performing two doublings.
First, let $M_1$ be the manifold obtained by gluing the tile $W^n$ in $M_0$ to $W^1$ in $M_0^*$ along $(n,1)$.
In other words, we start by joining $M_0$ and $M_0^*$ in a different way.
Then, take a CPT-transformed copy $M_1^*$ and glue it to $M_1$ along all the remaining $(n,1)$ geodesics as before.
The final result is a larger, redundant manifold on which there are twice as many RT surfaces in correspondence with a given $E_W$ surface in comparison to the case when $n$ is even.
The only exception to this count is for those curves with endpoints on $(n,1)^n$ in $M_0$ and $(n,1)^1$ in $M_0^*$, which become joined when $M_1$ is formed (and hence have doubled length compared to when $n$ is even).

\subsection{Higher dimensions}

Our results generalize straightforwardly to higher dimensions.
For $d>2$, the boundary subregions do not in general inherit a unique ordering from the topology of the boundary manifold, but we can choose such an ordering arbitrarily.
For any such choice, we can construct the replicated geometry $M_0$---and the final doubling to $M$---via the algorithm described in \Sec{sec:2d}.
As before, for simplicity we require either a static geometry or one that possesses a time-symmetric Cauchy slice, so that we can apply RT rather than HRT.

\begin{figure}[t]
    \centering
    \includegraphics[width=8.3cm]{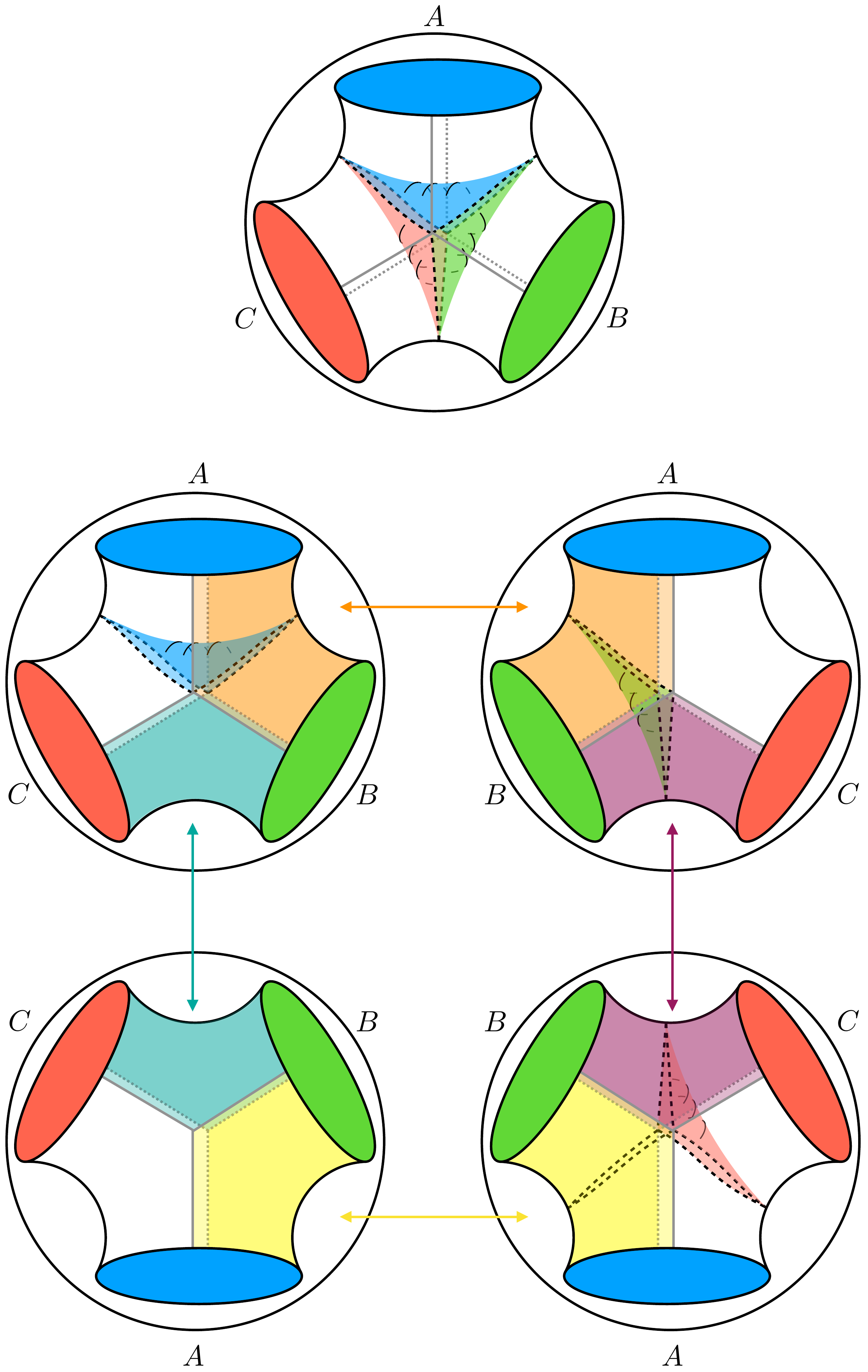}
    \caption{Formation of $M_0$ from four copies of a tripartite entanglement wedge in three (spatial) dimensions. Copies of the wedge are joined along shaded parts of $m(ABC)$ that have the same color.}
    \label{fig:3dconstruction}
\end{figure}

Instead of geodesics, we glue along the minimal codimension-two surfaces (codimension-one within a spatial slice) that are their analogues in the higher-dimensional geometry, forming a partition of the boundary of the entanglement wedge, as depicted in \Fig{fig:partitionD}.
To characterize this partition, first split each $\partial A_i$ into two simply-connected components $a_i$ and $b_i$.
The partition then consists of $n$ pieces $\pi_{1,2}$, $\pi_{2,3}$, \dots, $\pi_{n,1}$ such that, for each $1 \leq i \leq n$, $\pi_{i,i+1} \supset a_i, b_{i+1}$, $\pi_{i,i+1} \not\supset a_j, b_{j+1}$ for $j \neq i$, and $\cup_{i=1}^n \pi_{i,i+1} = m(A_1 \cdots A_n)$ (with all indices defined mod $n$ so that $n+1\equiv 1$).
The pieces $\pi_{i,i+1}$ are the analogues of the geodesics $(i,i+1)$ in \Sec{sec:2d}; see \Fig{fig:3dconstruction}.
This gluing can be accomplished  consistently with the junction conditions since these surfaces are extremal, as shown in detail in Refs.~\cite{Engelhardt_2019,dutta2019canonical}.
No shock wave in the energy-momentum tensor occurs at the junction (though gravitational wave shocks in the Weyl tensor can appear in the nonstatic case if the null geodesic congruence launched from the surface has nonzero shear).

All of the inequalities among $E_W$ surfaces derivable from our construction---e.g., those we obtained in \Sec{sec:ineqs}---carry over to the higher-dimensional case.
However, these relations are more powerful in higher dimensions, since they hold under any chosen ordering of the boundary regions.
There may be other constructions possible, beyond that presented in \Sec{sec:2d}, that apply for $d>2$ but not for $d=2$.
Such constructions could conceivably allow for the derivation of yet more exotic inequalities among the $E_W$ surfaces, but we leave the investigation of this possibility to future work.

\section{Discussion}
\label{sec:discussion}

Given a $n$-party entanglement wedge $W$, we described how to construct a larger manifold out of copies of $W$ and its CPT conjugate on which the areas of certain RT surfaces compute (integer multiples of) the areas of multipartite cross sections of $W$.
Entanglement entropy inequalities on the final manifold then translate to novel $E_W$ inequalities among entanglement wedge cross sections with nonuniform party numbers.
For a fixed ordering of $W$'s boundary regions, the construction gives the $E_W$ surfaces for all $n$-partite and lower ordered partitions of the boundary regions, cf. \Eq{eq:partition}.
The geometric aspects of our construction work in arbitrary numbers of dimensions; however, the correspondence between $E_W$ and entropic quantities in the boundary in arbitrary dimension is less fully understood.
While we presented our construction assuming that $W$ was a time-symmetric Cauchy surface or part of a static geometry, there do not appear to be obstructions in principle to extending it to dynamical settings, although the details and boundary entropic interpretation remain to be elucidated. 

This particular construction was originally motivated by the replicated construction given by one of the authors in Ref.~\cite{bao2019multipartite}.
This latter construction similarly computes the $n$-party $E_W$ as an entanglement entropy on a larger manifold.
It is simpler in that it requires fewer copies of $W$; however, it is only designed to obtain the $n$-party $E_W$.
In comparison, the construction described in the present work can be thought of as an extension that uses a minimal number of additional copies of $W$ to obtain a large class of $E_W$ surfaces with varying party number as RT surfaces on a single final manifold.

Other constructions that use different numbers of copies of $W$ and that have different gluing schemes are of course possible as well.
For example, the topological proposal of Ref.~\cite{Harper:2020wad} contains another such construction that computes the $n$-party $E_W$.
In terms of the language of \Sec{sec:general}, the construction in Ref.~\cite{Harper:2020wad} uses only the unprimed copies of $W$, which are then directly glued together to form a manifold without bulk geodesic boundaries instead of introducing a CPT double ($M_0^*$).
The construction is elegant in that is uses a minimal number of copies of $W$, but most $E_W$ surfaces aside from the $n$-party case do not map onto RT surfaces, or at least not minimal surfaces without self-intersections.

Based on the above observations, it is plausible that there exist other schemes that could overcome the ordering constraint on the $E_W$ surfaces that are simultaneously achievable with the construction described in this work.
These schemes could, e.g., use more copies of $W$, leave more bulk geodesics as open boundaries prior to doubling, etc.
The investigation of more elaborate constructions is a possible line of future research.

Another potential area of investigation is to study whether replacing $E_W$ in these inequalities with the various candidate boundary duals of $E_W$ can result in inequalities that can be proven for all quantum states; this a potential method for differentiating between these candidate boundary dual quantities.

A final interesting direction to pursue is to construct a holographic entropy cone that combines the entanglement entropy inequalities \cite{bao2015holographic, cuenca2019holographic} with the $E_W$ inequalities.
This is in particular motivated by so-called ``mixed'' inequalities, involving both $E_W$ quantities and normal entanglement entropies, suggesting that such a cone is not a simple tensor product of an $E_W$ cone and an entanglement entropy cone. This new cone would therefore be strictly more descriptive than the original holographic entropy cone and could lead to novel insights about the forms of entanglement permitted in holographic states.

\vspace{5mm}
\begin{center} 
{\bf Acknowledgments}
\end{center}
\noindent 
We thank Newton Cheng and Jonathan Harper for discussions.
N.B. is supported by the Department of Energy under grant number DE-SC0019380 and by the Computational Science Initiative at Brookhaven National Laboratory.
A.C.D. was supported for a portion of this work as a postdoctoral fellow (Fundamental Research) of the Research Foundation -- Flanders (FWO), [file number 12ZL920N]. / A.C.D. werd voor dit werk gedeeltelijk ondersteund door het Fonds Wetenschappelijk Onderzoek  -- Vlaanderen (FWO) als postdoctoraal mandaathouder, [dossiernummer 12ZL920N].
A.C.D. acknowledges the support of the Natural Sciences and Engineering Research Council of Canada (NSERC), [funding reference number PDF-545750-2020]. / La contribution d'A.C.D. {\`a} cette recherche a {\'e}t{\'e} financ{\'e}e en partie par le Conseil de recherches en sciences naturelles et en g{\'e}nie du Canada (CRSNG), [num{\'e}ro de r{\'e}f{\'e}rence PDF-545750-2020].
G.N.R. is supported at the Kavli Institute for Theoretical Physics by the National Science Foundation under Grant No. NSF PHY-1748958 and at the University of California, Santa Barbara by the Fundamental Physics Fellowship.

\bibliographystyle{utphys-modified}
\bibliography{references.bib}

\end{document}